\documentstyle[11pt,epsf]{article}
\def\ba{\begin{eqnarray}\samepage}
\def\ea{\end{eqnarray}}
\input amssym.def
\input amssym.tex

\def\double{\Bbb}

\newcommand{\po}{Poincar\'{e} }

\def\rr{{\double R}}
\def\zz{{\double Z}}

\newcommand{\mod}{\mbox{ mod\,}}

\newcommand{\diff}{\partial}

\newcommand{\be}{\begin{equation}}
\newcommand{\ee}{\end{equation}}
\newcommand{\ben}{\begin{eqnarray}\displaystyle}
\newcommand{\een}{\end{eqnarray}}

\def\Bbb#1{\fam\msbfam#1}

\def\expct#1{\langle #1 \rangle}
\def\beq{\begin{equation}}
\def\eeq{\end{equation}}

\begin{document}
\bibliographystyle{unsrt}

\title{The twelve dimensional super $(2+2)$-brane}
\author{Stephen Hewson\footnote{sfh10@damtp.cam.ac.uk}\,\, and
Malcolm Perry\footnote{malcolm@damtp.cam.ac.uk}\\ 
Department of Applied Mathematics and Theoretical
Physics\\Silver Street, Cambridge, CB3 9EW\\England}
\maketitle

\begin{abstract}

We discuss supersymmetry in twelve dimensions and present a covariant
supersymmetric action for a brane with worldsheet signature $(2,2)$,
called a super $(2+2)$-brane, 
 propagating in the
osp(64,12) superspace.  This
superspace is explicitly constructed, and 
is trivial in the sense that the spinorial part is a trivial bundle
over spacetime, unlike the twisted superspace of usual \po
supersymmetry. For consistency, it is necessary to take a projection
of the superspace. This is the same as the projection required for
worldvolume supersymmetry. Upon compactification of this
superspace, a torsion is naturally introduced and we produce the
membrane and type IIB string actions in 11 and 10 dimensional Minkowski
spacetimes. In addition, the compactification of the twelve
dimensional supersymmetry
algebra produces the correct algebras for these theories, including
central charges. 
These considerations thus give the
type IIB string  and $M$-theory  a single twelve dimensional origin.
\end{abstract}

\section{Introduction}
\setcounter{equation}{0}

The are five apparently distinct
known critical superstring theories in ten spacetime dimensions.
They are
\newcounter{xxx}
\begin{list}
{(\roman{xxx})}{\usecounter{xxx}}
\item type IIA,
\item  type IIB,
\item  $E_8\otimes E_8$ heterotic, 
\item   ${Spin(32)/ Z_2}$ heterotic,
\item   type I.
\end{list}
M-theory is a conjectured direct precursor of (i) and (iii), and
reproduces (ii), (iv) and (v) upon compactification to nine
dimensions \cite{mtheory}. More precisely, in
ten dimensions the ${Spin(32)/ Z_2}$ heterotic and type I theories are
S dual to each other, whereas the type IIB string is self-dual. In
nine dimensions, the two heterotic string theories are equivalent under
T-duality, as are the two type II strings. In addition, we find that  M-theory
compactified on the interval $I$ and the circle $S^1$ give the
$E_8\otimes E_8$ heterotic and IIA string theories respectively. 
Thus it is M-theory which provides the final link between the five
string theories. 
The mysterious M-theory, which is
supposed to exist 
in a spacetime of eleven dimensions,
ought somehow to be related to 
supergravity, and both two-branes and five-branes with $N=1$ worldvolume
supersymmetry, \cite{Witten: }. Some time ago, it was realized that
it was possible to construct type IIA ten-dimensional supergravity by the
dimensional reduction of eleven dimensional theory. Subsequently 
 Duff, Howe, Inami and Stelle,
\cite{DHIS:eleven} showed that the theory coming from the 
double dimensional reduction of the classical
two-brane from eleven to ten dimensions resulted in the classical
representation of the type IIA
superstring. It is natural therefore to ask if such a correspondence
could be made quantum mechanical. In string theory, the string coupling
constant $\kappa$ is given by $\kappa=e^{\expct\phi}$ where
 $\phi$ is the dilaton
field. The eleven dimensional interpretation of $\kappa$ requires it to be
identified with a component of the eleven dimensional metric. Suppose
that $g_{ab}$ is some  eleven dimensional metric on spacetime
${\cal{M}}$, and  there is a Killing
vector $k={\partial/\partial x^{11}}$ with a circle
 action on ${\cal{M}}$.  Identification of $x^{11}$ with unit period
then means that the radius of the Kaluza-Klein circle, $R$, is
determined by $g_{11,11}=4\pi^2 R^2$. Comparison of the two supergravity
theories then yields
\beq g_{11,11} =   e^{{4\over 3}\expct\phi}\,.  \eeq
Hence, the string coupling is related to the radius of the Kaluza-Klein
by
\beq R \sim \kappa^{2/3} \,.\eeq
Hence weak coupling string theory corresponds to $R$ small, whereas
infinite  coupling corresponds to flat eleven dimensional spacetime.

M-theory is hypothetical the eleven dimensional master theory that provides a
complete quantum mechanical version of this picture.
Compactification of M-theory on a circle
gives rise to type IIA string theory. Horava and Witten
subsequently showed that compactification of M-theory on ${S^1/Z_2}$
gives rise to the heterotic string with gauge group $E_8\otimes E_8$, 
\cite{HW:}.   A route to M-theory  therefore is to  explore
the eleven dimensional theories which are known to exist.

Effectively two string theories are omitted from this direct
unification. They are the 
heterotic string with $Spin(32)/Z_2$ gauge group, which is equivalent
to the type $I$ string in ten dimensions,  and the
type IIB string. 
As has been suggested by Vafa, \cite{Vafa:ftheory}, a natural
viewpoint would be to describe the type IIB string, that has
an $SL(2,\Bbb Z)$ S-duality, as arising from some kind of theory,
called F-theory, 
in twelve dimensions. Given a
manifold $A$ which is a $T^2$ fibration 
over a manifold $B$, F-theory on $A$ is defined to be equivalent to
the  type IIB
string on $B$. It is suggested that by allowing the RR
dilaton and axion to vary 
on the internal manifold we can explain the $SL(2,\Bbb Z)$ invariance
of the IIB string.  The implications of F-theory have been studied by
several authors \cite{ftheory}. 

The idea of there being twelve dimensions is not a new one: 
Supersymmetric $p$-branes moving in
dimensions higher than eleven have been discussed previously, most
notably by Duff and Blencowe
\cite{t6}, but not in any great detail  due to the
problems with a straightforward generalisation of 
the Green-Schwarz strings and membranes (An abortive attempt
 to 
 construct supersymmetric branes in higher dimensions was made
 by Blencowe, Duff, Hull and 
 Stelle in 1988 \cite{bdhs}).
A twelve
dimensional connection with the 
supergravity theories for $D\le 11$ has  been suggested by Hull
\cite{hull} and the possibility of a full twelve dimensional
supergravity theory has 
been studied previously \cite{twelve}. More recently we have seen the
appearance of several 
 other theories which make use of
a twelfth  
dimension \cite{twelfth,stheory,km,kmo}. It is now becoming apparent that
twelve dimensions 
could well 
have a role to play in the formulation of string and membrane
theories.  In order to explore this idea more fully it seems
necessary to try to find which   fundamental extended objects  can exist
in twelve dimensional spaces. It is the issue of twelve
dimensional fundamental $p$-branes which we address in this
paper. The results  will hopefully shed some light on the possible
true  nature
of a twelfth dimension.

\section{Supersymmetric Branes}
\setcounter{equation}{0}

We now consider the formulation of  supersymmetric theories in
 spacetimes of various dimension and signature. The goal is to 
discover consistent supersymmetric embeddings of $p$-branes into
twelve dimensional spacetimes. In this paper we will define a
$(p+q)$-brane to be a surface of signature  $(s,t)=(p,q)$. A $p$-brane
 will be a surface of signature $(s,t)$ with $s+t-1=p$. We shall use
 the word `spacetime' 
 to refer to any 
background manifold in which the $p$-brane exists. Since a
 supersymmetric theory requires the inclusion of fermions, we 
begin with a general discussion of spinors. 

\subsection{Spinors}

We start from  the Clifford algebra in $D$ dimensions

\be
\{{\Gamma_\mu,\Gamma_\nu\}^\alpha}_\beta=2\eta_{\mu\nu}{I^\alpha}_\beta\,
\ee
where $\eta=\mbox{diag}(-,-,\dots,-,+,+,\dots,+)$ is the metric on
the tangent space of
dimension $D=S+T$ and signature $S-T$. Thus there are $T$ timelike
directions, each corresponding to a minus sign in the metric, and  $S$
spacelike dimensions. We restrict ourselves to the cases $S\geq T$.  All the results we 
obtain are representation independent. We may always choose a
representation of the $\Gamma$ matrices such
that $\Gamma^\mu$ is hermitian (anti-hermitian) for $\mu$ spacelike
(timelike)  respectively. The indices $\mu$ and $\nu$ run
from $1\dots D$ and $\alpha$ and $\beta$ are spinorial indices. The spin
space has $2^{{int}(D/2)}$ dimensions, where `int' denotes the 
integer part.
If $S-T$ is even,  there are two possible inequivalent choices of
charge conjugation matrix, whereas if $S-T$ is odd there is a unique
charge conjugation matrix. 
In either case, the charge conjugation matrices  $C_\pm$ must  satisfy
\be
\tilde{\Gamma}_\mu=\pm C_\pm\Gamma_\mu C^{-1}_\pm \,,
\ee
where the tilde denotes transpose.
If we define $A=\Gamma^1\dots\Gamma^T$, where $\Gamma^\mu$ is timelike for
$1\leq \mu\leq T$,  then we have the relationship

\be
\Gamma^{\mu\dag}=(-1)^{T}A\Gamma^\mu A^{-1}\,,
\ee
and so we may define the  Dirac
conjugate as
\be
\bar{\psi}_D=\psi^{\dag}A\,.
\ee
This is chosen so that $\bar{\psi}_D\psi$ transforms as a scalar
under $SO(S,T)$ Lorentz transformations. The Majorana conjugate is defined as
\be
\bar{\psi}_M=\widetilde\psi C\,,
\ee
and a  Majorana spinor is one for which
$\bar{\psi}_D=\bar{\psi}_M$, corresponding to a real section of the
spin bundle. Clearly, both
these two spinors must satisfy the same Dirac equation, which leads to
 consistency conditions on the $C_\pm$, namely that Majorana spinors exist iff
there exists a  $C_+$ or $C_-$   such that \cite{nieuwen:leshouches}
\ben\label{maj}
\tilde{C}_+&=&(-)^{T+{int}((T+1)/2)}C_+\nonumber \\
\tilde{C}_-&=&(-)^{{int}((T+1)/2)}C_-\,.
\een
Since $\pm \Gamma^{\mu *}$ form equivalent representations of the
Clifford algebra, we may make a transformation from one representation
to the other by 
\ben
\Gamma^\mu&=&\eta B^{-1}\Gamma^{\mu *}B, \hskip1cm \eta=\pm 1\nonumber \\
\Rightarrow B^*B&=&\epsilon, \hskip1cm \epsilon=\pm 1\,,
\een
so that we can write the charge conjugation matrix as $C=\tilde{B}A$
\cite{kt}. The properties of the matrices $A$ and $B$ then imply that
\ben\label{gammaid}
{\tilde{\Gamma}}_\mu&=&(-1)^{T}\eta C\Gamma_\mu C^{-1}\nonumber \\
C^{\dag}C&=&1\nonumber \\
\tilde{C}&=&\epsilon\eta^T(-1)^{T(T+1)/2}C\,.
\een
 The possible choices of the numbers
$\eta, \epsilon=\pm 1$ 
depend on the signature of the spacetime as follows
\be\label{eta}
\begin{array}{|c|c|c|} \hline 
\epsilon &\eta&(S-T)\mbox{ mod}\,8\\ \hline
+1&+1&0,1,2\\ \hline
+1&-1&0,6,7\\ \hline
-1&+1&4,5,6\\ \hline
-1&-1&2,3,4\\ \hline
\end{array}
\ee
\medskip

We see that both $C_+$ and $C_-$ can be defined if $S-T$ is even,
otherwise only one of them exists. The charge conjugation matrix and
its inverse are then used to lower and raise spinor indices
respectively,  so that 
$\psi^\alpha=\psi_\beta (C^{-1})^{\beta\alpha}$ and
$\psi_\alpha=\psi^\beta C_{\beta\alpha}$.

The set of matrices
$\Gamma^{(i)\alpha\beta}=(\Gamma_{[\mu_1\dots\mu_i]}C^{-1})^{\alpha\beta}$,
where $i$ runs
over  $1\dots D$, and the square brackets denote antisymmetrisation,   
form a basis for the space of $2^{[D/2]}\times 2^{[D/2]}$ matrices. The
$(\Gamma^{(i)})^{\alpha\beta}$ 
are each either symmetric or antisymmetric in their spinor indices. By
using the   relations (\ref{gammaid}) we find that the parity, $\pi$,  of these
matrices is given by 
\be\label{parity}
\pi=\epsilon\eta^T(-1)^{\frac{T(T+1)}{2}}\left((-1)^T\eta\right)^i(-1)^{\frac{i(i-1)}{2}}\,.
\ee
From the table  (\ref{eta}) we can find a basis of 
matrices of given symmetry for a given definition of the $\Gamma$ matrices.

\subsection{$N=1$ superalgebras}

Superalgebras play an important role in the formation of $p$-brane
theories. A natural supposition is that it should be possible to
formulate the theory in flat spacetime in such a way that spacetime
supersymmetry is manifest. This is the rationale for the Green-Schwarz
approach to the theory of extended objects. 

Flat spacetime has the \po group IO$(S,T)$ as its isometry group. Locally this
may be described by the exponential of the Lie algebra generated by
the Lorentz rotations $M_{\mu\nu}$ and 
the translations $P_{\mu}$
\ben\label{poincare}
[M_{\mu\nu},M_{\rho\sigma}]&=& 
M_{\nu\sigma}\eta_{\mu\rho} +M_{\mu\rho}\eta_{\nu\sigma}-M_{\nu\rho}\eta_{\sigma\mu}- M_{\sigma\mu}\eta_{\nu\rho} \nonumber \\
\left[M_{\mu\nu},P_\rho\right]&=&P_\mu\eta_{\nu\rho}-P_\nu\eta_{\mu\rho}\nonumber \\
\left[P_\mu,P_\nu\right]&=&0\,.
\een

To implement a simple, $N=1$, spacetime supersymmetry we consider 
graded algebras generated by $\mbox{\{}M,P,Q\mbox{\}}$, where $Q$ is a
spinorial Grassman odd generator. Usually the term `supersymmetry algebra' refers to
the following extension of the bosonic algebra 
\ben\label{superpo}
\left[M_{\mu\nu},Q^{\alpha}\right]&=&-\frac{1}{2}{(\Gamma_{\mu\nu})^\alpha}_\beta
Q^\beta \nonumber\\
\left[P_\mu,Q\right]&=&0 \nonumber\\
\{Q^\alpha,Q^\beta\}&=&(\Gamma_\mu)^{\alpha\beta}P^\mu\,.
\een
There are, however,  many possible spinorial extensions of the \po algebra, and each will
generate some supergroup which may be used to define a spacetime
supersymmetry. However, only some of these will correspond to on-shell
supersymmetric theories.  Nahm \cite{nahm} investigated  supersymmetry
algebras satisfying the on-shell condition and restrictions on the
spins of the states in spacetimes with
Minkowski signature, leading to constraints on possible background
spacetimes.  We shall initially consider general
supersymmetric theories without concerning ourselves with  degrees of freedom or
signature. The restrictions which arise  come purely from the self
consistency of the superalgebras, by which we mean that the
super-Jacobi identities must be satisfied

\be
\mbox{[}A,\mbox{[}B,C\mbox{\}}\mbox{\}}=\mbox{[}\mbox{[}A,B\mbox{\}},C\mbox{\}}-(-)^{bc}\mbox{[}\mbox{[}A,C\mbox{\}},B\mbox{\}}\,,
\ee

with
\ben
(-)^{bc}&=&-1 \ \ \ \mbox{ if } B \mbox{ and }C \mbox{ are fermionic,}\nonumber \\
&=&+1 \ \ \ \mbox{otherwise.}\,
\een
 In addition to this constraint, we shall only require
that the graded algebra  reduce to the bosonic \po algebra, (\ref{poincare}), when we set
$Q=0$.
 To discover the possible extensions, we first write down  the
transformation law of a 
spinor under a  Lorentz transformation. This gives the first
expression in (\ref{superpo}). 

We now note that the  superalgebra must contain a
$\mbox{\{}Q,Q\mbox{\}}$ term.  We need
to determine the form of this term. Since the anticommutator 
must be symmetric in its spinor indices, we expand in terms of a
basis for symmetric matrices, which may be found from the expression (\ref{parity})
\be\label{s}
\mbox{\{}Q^\alpha,Q^\beta\mbox{\}}=\sum_k \frac{1}{k!}(\Gamma_{\mu_1..\mu_k})^{\alpha\beta} (Z^{(k)})^{\mu_1..\mu_k}\,,
\ee
where $k$ runs
over the values for which $\Gamma^{(k)}$ is symmetric in its
spinor indices. This is the approach discussed in \cite{t3,stheory}. We must then check that all possible  super-Jacobi
identities hold. This places restrictions on the $Z^{(i)}$ terms. Note
that the terms $Z^{(1)},Z^{(2)}$ have the same degrees
of freedom as the
generators $P,M$ respectively. Thus, if the term $\Gamma^{(1)}C^{-1}$
is antisymmetric in the spinor indices then the anticommutator of the
$Q$ with itself  cannot
generate momentum, and the theory cannot exhibit spacetime
supersymmetry in the usual sense of  (\ref{superpo}). The symmetry
of the matrix $\Gamma^{(1)}C^{-1}$ depends on the possible choice of the charge
conjugation matrix $C$. In twelve
dimensions  we may generally choose the charge conjugation matrix to be
either $C_+$ or $C_-$. For $C_+$ the
symmetric matrices are given by those $\Gamma^{(i)}C^{-1}$ for which
$i=2,3 \mod 4$. For $C_-$,  they 
occur for the  cases $i=1,2 \mod 4$. This result is independent of the
signature of the twelve dimensional space. Thus we find that in twelve
dimensions we may formulate a standard theory of supersymmetry for the choice
$C_-$ but not for the choice $C_+$: only for the $C_-$ case  may we extend
the \po algebra  by the terms (\ref{superpo}) if we identify $Z_\mu$
with $P_\mu$. For a Dirac spinor this result is independent of
$S-T$.  Imposing the              Majorana condition on
the spinors  leads to restrictions on the possible signature, which
are 
shown in the table for the case $S+T=12$
\be\label{table2}
\begin{array}{|c|c|c|c|}
 \hline
S&T&(\epsilon,\eta) \mbox{ for }C_+&(\epsilon,\eta) \mbox{ for }C_-\\
\hline
12&0&-&-\\ \hline
11&1&-&(1,1)\\ \hline
10&2&(1,1)&(1,-1)\\ \hline
9&3&(1,-1)&-\\ \hline
8&4&-&-\\ \hline
7&5&-&(1,1)\\ \hline
6&6&(1,1)&(1,-1)\\ \hline
\end{array}
\ee

The general extension of the \po algebra  involves the addition of
$Z^{(k)}$ terms other than $Z^{(1)}\sim P$. These $p$-form charges
\cite{t:pbranedem}  are   in some respects similar to  central charges
which occur only for  $N\geq 2$
supersymmetric theories, but in contrast arise from the commutator of the
spinor generator  with
itself. The
question may be asked as to which terms, if any,  arise naturally. To answer
this,  we
investigate the $p$-form charges  in the context of gradings of the de Sitter
algebra, given by 
\ben\label{desitter}
[M_{\mu\nu},M_{\rho\sigma}]&=& 
M_{\nu\sigma}\eta_{\mu\rho} +M_{\mu\rho}\eta_{\nu\sigma}-M_{\nu\rho}\eta_{\sigma\mu}- M_{\sigma\mu}\eta_{\nu\rho} \nonumber \\
\left[M_{\mu\nu},P_\rho\right]&=&P_\mu\eta_{\nu\rho}-P_\nu\eta_{\mu\rho}\nonumber \\
\left[P_\mu,P_\nu\right]&=&mM_{\mu\nu}\,.
\een
where $m^{-1}$ is  the radius of the de Sitter space. The de Sitter
algebra  is of
interest because it reduces to the \po algebra in the limit
$m\rightarrow 0$. If $m\neq 0$ then we find that the extra structure
leads to restrictions on the possible antisymmetric tensors. Since branes
are microscopic objects they are insensitive to large scale
structure. It is therefore natural to preserve these restrictions even
in the $m=0$ limit. 

The general
grading of the de Sitter algebra includes an anticommutator of the form
\be\label{generalq}
\{Q^\alpha,Q^\beta\}=\sum_{\mbox{\tiny{symmetric }}\Gamma}\frac{1}{k!}(\Gamma_{\mu_1\dots\mu_k})^{\alpha\beta}Z^{\mu_1\dots\mu_k}\,
\ee

We may choose any set of the $Z$ fields provided that they are 
consistent with the superalgebra. The restrictions which arise from
the super-Jacobi 
identities are given in \cite{t3} for the case of $C_-$, although a
similar analysis may also be done for the case $C_+$. We find that there
are only two consistent super-de Sitter algebras, for each $C$,
without setting to zero some 
of the possible  $Z^{(i)}$ in the $\{Q,Q\}$ anticommutator.    
We now impose the additional constraint that in the infinite radius
limit we reproduce the \po algebra structure.  Since $P$ and $M$ have the same degrees
of freedom as $Z^{(1)}$ and $Z^{(2)}$ respectively we should identify
them in the limit. Any other  $Z^{(i)}$ which remain in the $m=0$
limit will be taken to be $p$-form charges.  Consistency of the limit of
the de Sitter algebra 
with the \po case  rules out one of the aforementioned  solutions,   
leaving us with a single possible maximal algebra for each $C$. For
the $C_-$ case,  
the algebra is generated by the set $\{Q,P,M,Z^{(5)},Z^{(6)},Z^{(9)},Z^{(10)}\}$. The
only subalgebra of this algebra  is simply
$\{Q,P,M\}$  \footnote{Details of this calculation are given in
appendix A}. Thus, for the  algebra which generates momentum,  no $Z^{(i)}$
terms, for $i>2$,  are singled out from the full  set of $p$-forms.
 For the
$C_+$ case we  find the full  algebra
is generated by the set of generators
$\{Q,M,Z^{(3)},Z^{(6)},Z^{(7)},Z^{(10)},Z^{(11)}\}$. In this case, the algebra
contains only the
subalgebra  generated by $\{Q,M,Z^{(6)},Z^{(10)}\}$. We may suppose that we
make the identification
$M_{\mu\nu}=\frac{1}{10!}\epsilon^{\rho_1\dots\rho_{10}\mu\nu}Z_{\rho_1\dots\rho_{10}}$,
where $\epsilon$ is the alternating tensor.   
In the infinite radius limit for this truncation, after rescaling the
$Z^{(2)}$, we obtain  the
additional non-zero  commutation relations
\ben\label{subalgebra}
\left[M_{\sigma\rho},Z_{\mu_1\mu_2\mu_3\mu_4\mu_5\mu_6}\right]&=&\eta_{[\mu_1|\rho}Z_{\sigma|\mu_2\mu_3\mu_4\mu_5\mu_6]}-\eta_{[\mu_1|\sigma}Z_{\rho|\mu_2\mu_3\mu_4\mu_5\mu_6]}\nonumber
\\
\{Q^{\alpha},Q^{\beta}\}&=&\frac{1}{2}(\Gamma_{\mu\nu})^{\alpha\beta}M^{\mu\nu}+\frac{1}{6!}(\Gamma_{\mu_1\dots\mu_6})^{\alpha\beta}Z^{
\mu_1\dots\mu_6}\,.
\een
It is noteworthy that, after taking the limit,  it is possible to
include any number  of
the $Z^{(i)}$ into the algebra in a similar way to the antisymmetric
six index object in (\ref{subalgebra}), as  types of $p$-form  charges. The
above considerations 
for de Sitter space  show 
that only the above  cases of sets of generators which form subalgebras  arise
naturally if $P$ is excluded 
from the algebra. The only other natural truncation is simply the \po
algebra which, of course, generates momentum. It should be stressed
that we can only expect these algebras to be 
completely consistent for an on-shell theory. Off shell we inevitably
find anomalies in the super-Jacobi identities. For the algebra
including (\ref{subalgebra}) we find that all the super-Jacobi
identities are indeed satisfied on-shell. 

We now address the issue  of the degrees of freedom of the
spinorial part of the algebras. As a symmetric matrix in $2^{[D/2]}$
dimensions, $\{Q,Q\}$ has $\frac{1}{2}{.64.65}=2080$  components,
since 
the Majorana spinor has 64 real components. The two maximal sets of
generators $\{Z^{(k)}\}$,  for $k=\{1,2,5,6,9,10\}$ or
$\{2,3,6,7,10,11\}$,  both also have 2080 components, since each
antisymmetric tensor field has $\frac{12!}{(12-k)!k!}$
components. Thus a matching will  occur for the case of the maximal
algebra. For
any  truncation of one of the algebras this matching will not
occur. However, for the natural non-\po subalgebra which arises,
(\ref{subalgebra}), generated by $\{Q,M,Z^{(6)}\}$, we
may obtain  
saturation if we project out half of the spinor degrees of freedom. We
define  a pair of projection operators ${\cal{P}}_\pm=\frac{1}{2}(1\pm
X)$ where $X^2=1$, with $X$ chosen such that $X^2=1$ and so that 
${\cal{P}}_\pm$ are rank 
32.  Taking the projection of (\ref{subalgebra}) and the first two
terms in 
(\ref{superpo}), and  setting $Q_-=0$, we obtain
\ben\label{projalg}
[M_{\mu\nu},Q^\alpha_+]&=&-\frac{1}{2}{(\Gamma^+_{\mu\nu})^\alpha}_\beta Q^\beta_+\nonumber\\
{[P_\mu,Q^\alpha_+]}&=&0\nonumber \\
\{Q^{\alpha}_+,Q^{\beta}_+\}&=&\frac{1}{2}(\Gamma^+_{\mu\nu}
)^{\alpha\beta}M^{\mu\nu}+\frac{1}{6!}(\Gamma^{+}_{\mu_1\dots\mu_6})^{\alpha\beta}Z^{+
\mu_1\dots\mu_6}\,,
\een
where
$(\Gamma^+_{\mu_1\dots\mu_d})^{\alpha\beta}={\cal{P}}^\alpha_{+\gamma}{\cal{P}}^\beta_{+\delta}(\Gamma_{\mu_1\dots\mu_d})^{\gamma\delta}$.
The remaining part of the $Z^{(6)}$ term, $Z^{+(6)}$, is now self-dual
with respect to the 
projection operator, in that the complement of $Z^+$ in $Z$ vanishes
identically so that
$Z^{+\mu_1\dots\mu_6}\Gamma^-_{\mu_1\dots\mu_6}=0$.   The degrees of
freedom now match up:
the anticommutator  has
$\frac{32.33}{2}=528$ degrees of freedom; the self-dual six form and
the two form have 
$\frac{12!}{2.6!.6!}=462$ and $\frac{12.11}{2}=66$ degrees of
freedom respectively. Other than the maximal extension of the \po
algebra for each choice of $C$, this is the only algebra in which the
matching  of degrees of
freedom in the anticommutator term occurs. 
For completeness, we present an on-shell  representation of this 
algebra for which we treat the $Z^{(6)}$ term as central
\ben{\label{sio2rep}}
P_\mu&=&\frac{\diff}{\diff X^\mu}\,,\nonumber\\
Q_\alpha&=&\frac{\diff}{\diff\theta^\alpha}\,,\nonumber\\
M_{\mu\nu}&=&X_\mu\frac{\diff}{\diff X^\nu}-X_\nu\frac{\diff}{\diff
X^\mu}+\frac{1}{2}\theta_\alpha(\Gamma^{\mu\nu})^{\alpha\beta}\frac{\diff}{\diff\theta^\beta}\,.
\een

Note that if we define $X=i^T\Gamma^{13}$ in the expression for the
projectors ${\cal{P}}_\pm$, where
$\Gamma^{13}=\Gamma_1\dots\Gamma_{12}$, then ${\cal{P}}$ is the Weyl
projector, and self-dual means self-dual with respect to
the Hodge star operator. Other choices of projector are perfectly
possibly, although they will not be Lorentz invariant. This does not
imply  that the underlying theory containing such spinors is
not Lorentz invariant, however, which is an important point to note.
For example,  consider a twelve 
dimensional theory
in which arbitrary spinors $X,Y$ are constrained to satisfy the
identity
\be\label{lorentzcond}
X_\alpha(\Gamma_{\mu\nu})^{\alpha\beta}Y_\beta=0\hskip1cm \forall
\mu,\nu=1\dots 12\,.
\ee
This is a completely Lorentz invariant constraint, and hence describes
a Lorentz invariant theory. If we wish
to employ  an explicit representation of the spinors obtained by a
projection of 
Majorana spinors in $D$ dimensions,  then the projector used cannot be
Lorentz invariant if $T=2$, although it can if $T=1$. In some
signatures this behaviour may create
the  naive impression  that 
the underlying theory
described by (\ref{lorentzcond}) is not Lorentz invariant.

\bigskip

We have seen in this section  that  there are many
possible graded 
extensions of the \po
algebra, and so the term `super-\po algebra' is,  therefore, sometimes 
ambiguous. We shall extend the idea of a  super-\po algebra to 
sio$(S,T)_{\{k\}}$,  the algebra 
defined by (\ref{poincare}), the first two terms in (\ref{superpo}) and
\ben
\{Q,Q\}&=&\sum_{k=i_1\dots
i_n}\frac{1}{k!}\Gamma^{(k)}Z^{(k)}\,\nonumber\\
\left[Z_{\mu_1\dots\mu_i},M_{\rho\sigma}\right]&=&\eta_{[\mu_1|\rho}Z_{\sigma|\mu_2\dots\mu_i]}-\eta_{[\mu_1|\sigma}Z_{\rho|\mu_2\dots\mu_i]}\,\forall
i\nonumber\\
\left[Z^{(i)},Z^{(i)}\right]&=&0,\hskip1cm i,j\neq 2\,\nonumber \\
Z^{(1)}\equiv P& & Z^{(2)}\equiv M\,.
\een
Each of these superalgebras may correspond to a number of
supergroups. Since the bodies of these  algebras are themselves  Lie
algebras, the 
relationship between the superalgebras and the supergroups seems to be well
understood \cite{t4} and the theory progresses in close analogy with
that for   bosonic Lie algebras and Lie groups.  Our results will only
depend  local 
properties  and it will therefore  be consistent to discuss the corresponding
($N=1$) 
supergroup SIO$(S,T)_{\{k\}}$ in this work.  We can think
of the supergroup  as acting on some
superspace. A superspace is a
$\zz_2$-graded vector space consisting of a 
Grassmann even (bosonic)  subspace and a Grassmann odd
(fermionic) subspace. In a local region of superspace, the supergroup
elements  can be found via 
exponentiation of the superalgebra. Superspace is a
generalisation of 
the usual notion of spacetime.

\subsection{Super $p$-branes in an sio$(S,T)_1$ invariant background}

We now describe the formulation of a $p+1$ dimensional extended
object, or $p$-brane, 
moving in a flat $12$-dimensional spacetime background which is
invariant under the 
action of a super-\po group locally described by the algebra
sio$(S,T)_1$. We  suppose that there
are $s$-spacelike directions and $t$-timelike directions on the
brane so that  $s+t-p=1$. In order to allow a consistent embedding of
the brane into spacetime we
make the restrictions $s\leq S$ and $t\leq T$. It is natural to
formulate such a theory on the superspace associated with 
ISO$(S,T)_1$, which is described by the local coordinates

\be
Z^M=(X^\mu,\theta^\alpha)\,,
\ee
where the $X$ are the bosonic spacetime coordinates and the $\theta$
are anticommuting Majorana spinors.  
The action of a supergroup element infinitesimally close to the
identity gives the following change in the coordinates
\be\label{coordtrans}
\delta\theta^\alpha=\psi^\alpha \hskip1cm \delta
X^\mu=a\bar{\psi}\Gamma^\mu\theta\,
\ee
where $\psi$ is a constant Majorana spinor. The parameter $a$
is chosen so the variation in the bosonic coordinates is real. This
requires that 
\be
a^*=\epsilon\eta(-1)^{T}a\,,
\ee
where the values $\eta$ and $\epsilon$ are given in (\ref{eta}).  From
these relationships  we can define forms invariant under the action
of SIO$(S,T)_1$
\be
\Pi^\mu=dX^\mu-a\bar{\theta}\Gamma^\mu d\theta\hskip1cm
\Pi^\alpha=d\theta^\alpha\,.
\ee
We can then construct the action for the $p$-brane by considering the
pullback of these invariant forms to the worldvolume \cite{t2};  $dX\rightarrow
\diff_iXd\xi^i$,  where $\xi$ are  the coordinates on the
$p$-brane. The analogue of the Nambu-Goto action 
action obtained  from the principle of least action is  
\ben\label{brane}
S_0&=&\int
d^{p+1}\xi\,[\mbox{det}({\Pi^\mu}_i{\Pi^\nu}_j\eta_{\mu\nu})]^{\frac{1}{2}}\nonumber\\
{\Pi^\mu}_i&=&\diff_iX^\mu-a\bar\theta\Gamma^\mu\diff_i\theta\,,
\een
which may be re-written in first order form, using the Howe and Tucker
technique \cite{ht:action}, as
\be
S_0=\int
d^{p+1}\xi\sqrt{|g|}\left(\frac{1}{2}g^{ij}{\Pi^\mu}_i{\Pi^\nu}_j\eta_{\mu\nu}-\frac{1}{2}(p-1)\right)\,,
\ee
where $g^{ij}$ is an auxiliary metric on the $p$-brane. These actions
are  manifestly spacetime supersymmetric. We also require that
there be supersymmetry on the brane.  
This necessitates  a matching between the worldvolume bosonic and
fermionic degrees of freedom. Can such a matching occur on a brane in
twelve dimensions?

In theories of extended objects in non-Euclidean spacetimes,  negative
norm states exist due to propagation in timelike directions. To avoid
this problem  we may fix a gauge using
diffeomorphism invariance of the brane to allow only propagation in
directions transverse to the worldvolume. This gives us positive norm
states if the residual symmetry group $SO(S-s,T-t)$ is compact, which
requires that $T=t$. The gauge fixing leaves $D-p-1$ transverse coordinates,
which correspond to  the bosonic degrees of freedom. For a 
three-brane in twelve dimensions we have eight transverse $X^\mu$. This
is the important  case which parallels the string and the membrane in ten and
eleven dimensions respectively. 

We should now count the spinorial degrees of freedom. A Dirac 
spinor in twelve dimensions has $2^{[12/2]}=64$ complex components,
and a Majorana spinor has 64 real components. The Majorana spinor has
32 real on-shell degrees of freedom, and this is halved to 16
components by taking the Weyl projection, or any other rank 32
projection.   In order to obtain a matching for  the 8 bosonic coordinates we
need to find  
an extra fermionic symmetry. To find such a symmetry we must introduce  an
additional term in the action, called the Wess-Zumino term
\be\label{WZ}
S_{WZ}=-\int d^{p+1}\xi\left(\frac{2}{(p+1)!}\epsilon^{i_1\dots
i_{p+1}}B_{i_1\dots i_{p+1}}\right)\,
\ee
where $B_{i_1\dots i_p+1}$ are the components of a $p+1$ form $B$
which is the potential for the $p+2$ superspace form $H=dB$. This term
is of the correct form          since it is spacetime
supersymmetric and transforms in the same way as the action $S_0$
(\ref{brane}) under  a scaling of the
superspace coordinates. The full action $S_0+S_{WZ}$ is then
invariant \cite{t5,t6}, for  any choice of
signature,  under the so called $\kappa$-symmetry transformation
\be\label{ktrans}
\delta\theta=\frac{1}{2}(1+\Gamma)\kappa \hskip1.5cm \delta X^\mu =
\frac{a}{2}\bar{\theta}{\Gamma}^\mu(1+\Gamma)\kappa\,,
\ee
where  $\kappa$ is a scalar on the brane and a fermion in
spacetime. The matrix $\Gamma$ is essentially the Weyl projector on
the brane,  and is given by 
\be{\label{gamma}}
{\Gamma^\alpha}_\beta=\frac{(-1)^{(p+1)(p+2)/4}}{(p+1)!\sqrt{|g|}}\epsilon^{i_1\dots
i_{p+1}}E^{\mu_1}_{i_1}\dots
E^{\mu_{p+1}}_{i_{p+1}}{(\Gamma_{\mu_1\dots\mu_{p+1}})^\alpha}_\beta\,,
\ee
where $E^i_\mu$ is the vielbein \cite{t6}. $\Gamma$ has the property that
$\Gamma^2=1$ for certain values of the brane signature, hence 
$\frac{1}{2}(1\pm\Gamma)$ may be  projection operators, which may in fact
be used to gauge away half the spin degrees of freedom.   
By employing the $\kappa$-symmetry and making use of the equations of
motion we obtain the general  matching formula
\be
D-p-1=\frac{1}{4}N\,,
\ee
where $N$ is the dimension of the spin space, or its projection if
 the  restriction $\theta\rightarrow{\cal{P}}\theta$ is made on the
spinors. This equation may only be 
satisfied in twelve dimensions if we take $p=3$ and have eight
fermionic degrees of freedom. To count the spinor degrees of freedom
we first note that a $\kappa$-symmetry exists 
if the four  dimensional worldvolume may be chosen to be  Weyl, which
is the case for a brane of signature $(2,2)$. Imposing the effects of
the $\kappa$-symmetry and then imposing the equations of motion and
the Majorana condition 
leaves us with 16 degrees of freedom. 
To reduce this to eight, and hence obtain worldvolume supersymmetry,
we may act with 
the Weyl projector on the spacetime spinors. 
We may do this 
consistently if the
spacetime 
signature is $(10,2)$, since Majorana-Weyl spinors may only be defined
if $(S-T)\mod 8=0$ \cite{nieuwen:leshouches}. 
  
We shall now check the consistency of this construction.  
The $\kappa$-symmetry is defined via a Wess-Zumino action involving
the $(p+2)$-form $H$
\be\label{h}
H=\frac{a}{2p!}\Pi^{\mu_p}\dots\Pi^{\mu_1}d\bar{\theta}\Gamma_{\mu_1\dots\mu_p}d\theta\,.
\ee
For this to be well defined we require that
$H$-be non-zero and closed. Since the $d\theta$ are commuting
variables, for $H$ to be not identically zero, $\Gamma^{(p)}$ must be 
is symmetric in its spinor indices. If  $dH=0$ then we must have that 
\be\label{formeq}
({\cal{P}}d\bar{\theta}\Gamma_\mu
{\cal{P}}d\theta)({\cal{P}}d\bar{\theta}\Gamma^{\mu\mu_1\dots\mu_{p-1}}{\cal{P}}d\theta)=0\,,
\ee
which has solutions only if \cite{ach:matching}
\be
D-p-1=\frac{1}{4}N\,
\ee
This condition is  known to be sufficient if 
 $1\le D\le 11$, and it can be  shown\footnote{see appendix
A} that the  identity holds
for certain choices of projector for $D=12$
and $p=3$,  which  means that the Wess-Zumino term is well defined.
This being so it would appear that we could  define the $\kappa$-symmetry
and hence a super three-brane. However, we actually encounter an
inconsistency for the three-brane moving in an
SIO$(S,T)_1$ background: $\Gamma^{\mu}$ and
$\Gamma^{\mu_1\mu_2\mu_3}$ cannot be simultaneously symmetric since
$\Gamma^{(i)}$ and $\Gamma^{(i+2)}$ always have the opposite
parity. This implies that $H\equiv 0$ if the anticommutator of $Q$ with
itself  generates momentum,
which means that  the Wess-Zumino term is identically equal to
zero. This is a problem since the brane is expected to couple to the
local version of the  supersymmetry theory via the Wess-Zumino
form. In addition,  we may not  define a 
$\kappa$-symmetry either, which means that the degrees 
of freedom on the worldvolume do not 
match up and the brane is consequently ill defined.     
  
We have learnt that we may not formulate a three-brane
theory in twelve 
dimensions after all, if we assume invariance of the brane under the standard 
 supersymmetry group SIO$(S,T)_1$. This type of consistency problem does not
arise in ten or eleven dimensions for which we must choose $p=1,2$ 
respectively for the $H$-form. Since  we can make $\Gamma^{(1)}$ and
$\Gamma^{(2)}$ 
symmetric simultaneously, we encounter no problems in the formulation
of a string or membrane theory for the usual supersymmetry algebra.
We must conclude that 
in twelve dimensions the situation is more complex than it at first
appears. The problem may appear to be insoluble, since $\Gamma^{\mu}$
and $\Gamma^{\mu\nu\rho}$ are never simultaneously symmetric. There is
a resolution, however, which requires us to relax the notion of what
we mean by supersymmetry.

 In order to progress we need to carry out
the previous  
procedure  using the algebra sio$(S,T)_2$ which differs from
sio$(S,T)_1$ in that the $\{ Q,Q\}$ anticommutator in (\ref{superpo})
is replaced by  
\be\label{newcom}
\{Q^\alpha,Q^\beta\}=-\frac{1}{2}(\Gamma_{\mu\nu})^{\alpha\beta}M^{\mu\nu}\,.
\ee
This expression requires 
$\Gamma^{\mu\nu}$ to be symmetric, but there is no restriction  on
$\Gamma^\mu$. We may now define a non-trivial    superspace
form $H$  , since it is possible that $\Gamma^{\mu\nu}$ and
$\Gamma^{\mu\nu\rho}$ are simultaneously  symmetric. Trying to
formulate a theory in a new type of supersymmetry background may seem
strange, but there is no logical reason why this should not be done. 

We must now analyse the consequences of having an anticommutator of
the form (\ref{newcom}) in the supersymmetry algebra;  Although
sio$(S,T)_2$ reduces to the \po algebra in the limit where the 
supersymmetry generators are set to zero, the action on superspace will be
quite different since the commutator of the two supersymmetries will
generate an $SO(S,T)$ rotation instead of a momentum boost. Clearly we
shall  need to
carefully discuss the action of the group on the spacetime manifold.

\subsection{Action of the new supergroup SIO$(S,T)_2$ and the
$\mbox{sio}_2$ superspace}
Given a group $G$ with an invariant subgroup $H$,  we may define a coset
manifold
\be
{\cal{M}}=G/H\,,
\ee
where $G$ is the isometry group of the tangent space of the manifold and $H$ is the
isotropy subgroup for each  point, the group elements locally being
given by the 
exponentiation of the Lie algebra of $G$ \cite{t4}. Such a
construction may be used to 
generate homogeneous spaces.  As an example we note that Minkowski
space is given by the quotient of the \po group by the subgroup of
Lorentz rotations, $SO(S-1,1)$. Since $G$ is the isometry group of 
the tangent space of  ${\cal{M}}$, the action of an
infinitesimal group element leaves ${\cal{M}}$ invariant. From the
action of the group on the local coordinates of the manifold we can
find invariant one-forms from which an  action for the space may 
be constructed. We now consider such a procedure for the algebra sio$(S,T)_2$.
Since this algebra is  generated by $\{M,P,Q\}$,  a general group element
is given by
\be
G(X,\theta,\omega)=\exp(X^{\mu}P_\mu+\theta_\alpha
Q^\alpha+\frac{1}{2}\omega^{\mu\nu}M_{\mu\nu})\,,
\ee
where $(X,\theta,\omega)$ are parameters in the group space. As an ansatz we
shall suppose  that the background spacetime is given by the coset
\be\label{coset}
G(X,\theta,\omega)=\exp(X^{\mu}P_\mu+\theta_\alpha Q^\alpha)\exp(\frac{1}{2}\omega^{\mu\nu}M_{\mu\nu})\,,
\ee
in which case $(X,\theta)$ become the superspace coordinates. Since we
have replaced Minkowski space by a more general coset space, we should
question the choice of the subgroup we 
quotient by. By considering what happens in de Sitter space (appendix
A) we see that 
the subgroup generated by $M_{\mu\nu}$ is singled out
from those generated by the other $Z^{(i)}$, and it is thus natural to
write the quotient as in (\ref{coset}).  
We now act on the left of the coset with an infinitesimal group
element to give   
\be\label{product}
G(\delta X,\delta\theta,0)G(X,\theta,\omega)= \exp(\delta X^{\mu}P_\mu+\delta\theta_\alpha Q^\alpha)\exp(X^{\mu}P_\mu+\theta_\alpha Q^\alpha)\exp(\frac{1}{2}\omega^{\mu\nu}M_{\mu\nu})\,,
\ee
which  may be rearranged  using the Baker-Campbell-Hausdorff formula,
 given by
\be\label{bch}
\exp(\epsilon A)\exp(B)=\exp\left(B+\epsilon A +
\epsilon\sum_{n=1}^\infty\frac{1}{(n+1)!}[\underbrace{[\dots[A,B],B],\dots,B}_{\mbox{{n
times}}}]+O(\epsilon^2)\right)\,,
\ee
where $\epsilon$ is an infinitesimal superspace parameter. 
This expression  is greatly simplified for the sio$(S,T)_1$ case  because
only the first commutator in the series is non-zero. The series
consequently 
terminates,   giving rise  
to the supersymmetry transformations (\ref{coordtrans}). This
termination does not
necessarily occur for the sio$(S,T)_2$ algebra since  the commutator
of the $Q$ with 
itself  generates a rotation, which does not trivially commute
with any of the generators. We find that the right hand side  of
(\ref{product}) becomes 
\be
(\theta +\delta\theta)_\alpha Q^\alpha+(X+\delta X)^\mu
P_\mu +C_1+\sum_{n=2}^\infty \frac{1}{(n+1)!}C_n\,,
\ee
with 
\be
C_1=\frac{1}{2!}\delta\theta_\alpha(\Gamma^{\mu\nu})^{\alpha\beta}\theta_\beta
M_{\mu\nu}\,,
\ee
where $C_n$ is the commutator of $C_{n-1}$ with 
the exponent of the first factor of 
 $G$ in the expression (\ref{coset}).
In general this leads to an infinite 
series of terms  involving $M_{\mu\nu}$. These terms may not be
factored out into a coset 
form and as a result the construction is not generally  self-consistent. It
can be made so, however, if we make
the restriction that the variation in $\theta$ obeys the relationship
\be\label{constraint}
\delta\theta_\alpha(\Gamma^{\mu\nu})^{\alpha\beta}\theta_\beta
M_{\mu\nu}=0\,,
\ee
which corresponds to the vanishing of $C_1$ and hence all the other
$C_n$. 
The identity (\ref{constraint}) is not satisfied  for  general spinors
$\theta$;
to make (\ref{constraint}) hold, we restrict the form of $\theta$ so
that we are dealing with a subset of all possible spinors. The
projection onto this subspace is defined by a projection operator
${\cal{P}}$ so as to give 
\be\label{constraintp}
{{\cal{P}}_\alpha}^{\hat{\alpha}}\delta\theta_{\hat{\alpha}}(\Gamma^{\mu\nu})^{\alpha\beta}{{\cal{P}}_\beta}^{\hat{\beta}}\theta_{\hat{\beta}}\equiv
\delta\theta_\alpha\left(\widetilde{{\cal{P}}}\Gamma^{\mu\nu}C^{-1}{\cal{P}}\right)^{\alpha\beta}\theta_\beta=0\,,
\ee
 where
$\Gamma^{\mu\nu}\equiv{\{(\Gamma^{\mu\nu})^{\alpha}}_\beta\}$. For
this to be true for all variations requires us to restrict our spinors
so that
\be
\phi_\alpha(\Gamma^{\mu\nu})^{\alpha\beta}\psi_\beta=0\hskip0.5cm\forall
\phi,\psi\,.
\ee
This identity will prove to be fundamental in the following work and
is the defining feature of the $\mbox{sio}_2$ superspace. 
The
equation itself really defines a particular class of
spinors\footnote{The definition of these spinors is reminiscent of
that for 
pure spinors}, which
can, in the cases we are interested in, be obtained via a projection
of a general Dirac spinor.  Some
projectors  which satisfy this equation will be discussed in the
next section, within the context of compactification to lower
dimensions.   It is worth
noting that  the  equation  (\ref{constraintp}) is not satisfied in an even dimensional
space if we choose
${\cal{P}}$ to be the Weyl projector, unless there are an odd number
of timelike directions. Thus, for a (10,2) signature theory,  an
explicit projector used to obtain the 
spinors which satisfy (\ref{constraintp}) from a Dirac or Majorana
spinor will not be Lorentz invariant. In order to satisfy the
superspace identity we must in fact couple the Weyl
projector with another rank(1/2) projector, which is not a Lorentz
invariant procedure. The constraint itself is,
however, completely Lorentz invariant and the underlying theory
 may be fully covariant. 

As a corollary to this, if the identity (\ref{constraintp}) is
satisfied then  the supergroup action induces the very
simple $\mbox{sio}_2$ supersymmetry transformations
\be\label{suptrans}
\delta X=x \hskip1cm \delta\theta=\epsilon\,,
\ee
where $x$ is a constant commuting vector parameter and $\epsilon$ is
a constant anticommuting spinor parameter. This is very pleasing from a
geometrical point of view because the spinorial  part of the superspace is
now 
just a trivial bundle over spacetime, a superspace  generalisation of
$\rr^n$. This is, in a sense, a much more natural result then that
obtained for the usual sio$(S,T)_1$ algebra, (\ref{coordtrans}), in which the superspace is
twisted!

\subsection{ ${\mbox{sio}_2}$ $p$-branes}

Now that the new superspace has been defined we may construct actions
for $p$-branes propagating in such backgrounds.  The 1-forms
invariant under  the action of sio$(S,T)_2$ are simply given
by
\be\label{trivform}
\Pi^A=(dX^\mu,d\theta^\alpha)\,,
\ee
where $A$ runs over the even and odd coordinates. Using the pullback
of the forms
$\Pi^A$  to the brane, $\Pi_i^A=(\diff_iX^\mu,\diff_i\theta^\alpha)$,
we can write  down the superspacetime supersymmetric canonical action
\be{\label{superaction}}
S=\int
d^{p+1}\xi\left[\det(\Pi_i^A\Pi_j^B{\cal{G}}_{AB})\right]^{\frac{1}{2}}\,,
\ee
where the `metric' on the superspace is given by
\be
{\cal{G}}_{AB}=\pmatrix{\eta&0\cr 0&C}\,.
\ee
A brief discussion of  such metrics is given in
\cite{freund-kaplansky}. 

We are now in a position to repeat the  analysis for a three-brane
action  in an $\mbox{sio}_2$ background. Recall that the obstacle to
defining such an action in a 
usual $\mbox{sio}_1$ superspace background was
that the Wess Zumino
integral vanishes identically in twelve dimensions if the
supersymmetries generate  momentum, since $\Gamma^{(3)}$ is 
antisymmetric in such a situation. For $\Gamma^{(3)}$ 
to be symmetric we must necessarily choose  $C=C_+$. In such a case,
$\Gamma^{(2)}$ is symmetric, and may therefore appear on the right
hand side of the $\{Q,Q\}$ anticommutator. This means that
we may define a three-brane Wess-Zumino term if we employ the $\mbox{sio}_2$
superspace.

The new Wess-Zumino action is defined in  terms of the
trivial forms $\Pi^A$ given by (\ref{trivform}). Since there is no
torsion, we have  $d\Pi^A=0$ and the superspace form
$H=\Pi^\mu\Pi^\nu\Pi^\rho\Pi^\sigma
d\bar{\theta}\Gamma_{\mu\nu\rho\sigma}d\theta$ is automatically
closed. The underlying structure is now self consistent, so all that
remains to do is to  check  the brane degrees of freedom. 

 Constructing the Wess-Zumino action in the same way as for the
(\ref{WZ}) we find that 
\ben\label{wz2}
S_{WZ}&=&-\int *B\nonumber\\
B&=&dX^\mu dX^\nu dX^\rho
dX^\sigma d\bar{\theta}\Gamma_{\mu\nu\rho\sigma}\theta\,,
\een
where $*B$ is the pullback of the form $B$ to the worldvolume, and
$H=dB$. 
Since the $\mbox{sio}_2$ superspace is flat without torsion, there is
no spinorial part to the forms $\Pi^\mu$, hence there cannot be any
$\kappa$-symmetry in the sense of (\ref{ktrans}). If we insert a
general spinor transformation, then we find that under the variation
we obtain the non-vanishing term 
\be
\delta\theta_\alpha(\Gamma\Gamma_i)_{\alpha\beta}\theta=0\,,
\ee
where $\Gamma$ is defined in (\ref{gamma}) and $\Gamma_i$ is the
pullback of the $\Gamma^\mu$ to the brane. This term does not vanish
in general, and a $\kappa$-symmetry may not be used to gauge away
half the spinorial degrees of freedom. This is not a problem in the case for
which the superspace projector is not satisfied by the Weyl
projector. In these situations we need further project out another
half of  the
spinor coordinates to satisfy the supersymmetry constraint
(\ref{constraintp}). This extra projection has the same effect as the
$\kappa$-symmetry on the fermions and leads to  the existence of worldvolume
supersymmetry and hence the three-brane.  This is a very interesting
point: the existence of 
the worldvolume supersymmetry for the twelve dimensional three-brane,
and the consistency of the spacetime supersymmetry are inextricably
linked, whereas in the lower dimensional cases they could exist
independently.

\bigskip

To conclude this section we reiterate the main result. It is possible
to define a consistent supersymmetric three-brane action in twelve
spacetime dimensions. Such a theory has eight Bose and eight Fermi degrees of
freedom on 
the worldvolume and is thus a new member of the octonionic sequence \cite{kt}, in
addition to the ten dimensional superstring and the eleven dimensional
supermembrane. In order to define the Wess-Zumino term for the
three-brane it
is necessary 
to have $(\Gamma^{\mu\nu\rho})_{\alpha\beta}$ symmetric in the spinor
indices, which rules out the usual super-\po group, SIO$(S,T)_1$ as
the isometry 
group of the background spacetime, and also forces us to make the choice
$C=C_+$.    Taking an unusual  $N=1$
supersymmetric algebraic
extension of the \po algebra provides us with  a new supersymmetry
algebra in twelve 
dimensions, which we denote by sio$(S,T)_2$. The
action of the corresponding supergroup leads to a natural, trivial
superspace in which the brane propagates, if we make an
appropriate projection of the spinors. It is by such a projection that
we obtain  matching of
degrees of freedom, and hence supersymmetry,  on the brane. This
result gives a pleasing self consistency to the construction.

\section{Compactification from twelve dimensions}
\setcounter{equation}{0}
As we have shown in the previous section, it is possible to define a
consistent twelve dimensional supersymmetric three-brane theory. 
What relationship does this brane hold to the traditional string and
membrane? To answer this question we 
consider the simultaneous dimensional  reduction of the twelve
dimensional theory  
 described by the action (\ref{superaction}) with a
Wess-Zumino term. Our goal is to relate the twelve
dimensional 
theory to  the string
theories and M-theory, in ten and eleven dimensions of Minkowskian
signature respectively, by a compactification in which we remove one of
the additional timelike directions. Thus we can proceed in
three ways:  via a timelike reduction to eleven dimensions to give the
correct degrees of 
freedom for the $M$-theory two-brane; via a timelike reduction
followed by a spacelike reduction to obtain the ten dimensional IIA
string, and finally via  a double null reduction which
produces the ten dimensional IIB string.

Since the three-brane action may only be defined for the choice
$C=C_+$, 
we are restricted to consider theories with signature $(10,2)$,
$(6,6)$ or
$(9,3)$ (see table \ref{table2}). The case $(6,6)$ is excluded because
there are more timelike directions than can fit on the brane, and
hence the system is not classically stable.  We shall only consider  the case
$(S,T)=(10,2)$ in this paper, since this is the case which seems to be
directly related to the Minkowski theories in lower dimensions. We
shall henceforth refer to the three-brane as being, more precisely, a
$(2+2)$-brane.   For
simplicity we shall work with the
following real representation of Spin$(10,2)$:
\be\label{gamma12}
\Gamma_p=\pmatrix{\gamma_p&0\cr  0&-\gamma_p}\hskip0.5cm p=0\dots 9,\hskip1cm
\Gamma_{11}=\pmatrix{0&1\cr  1&0}\,,\hskip1cm
\Gamma_{12}=\pmatrix{0&1\cr  -1&0}\,,
\ee
where 
\be\label{clifford}
\{\Gamma_\mu,\Gamma_\nu\}=2\eta_{\mu\nu}\hskip1cm
\eta_{\mu\nu}=\cases{+1,&  if $\mu=\nu=1\dots 9$  or 11 \cr  -1,& if
$\mu=\nu=0$  or 12\cr  0,& otherwise.\cr }\,
\ee
The  $\gamma_p$ are a set of Spin$(9,1)$ gamma matrices, with $\gamma_0^2=-1$. We may choose a basis such that
the twelve dimensional spacelike (timelike) gamma matrices are
hermitian (anti-hermitian) respectively. We may also choose that 
\be\label{gamma11}
\gamma_{11}\equiv\gamma_0\dots\gamma_9=\pmatrix{I&0\cr 0&-I}\equiv J\,.
\ee
It may easily be shown that the twelve dimensional charge conjugation
matrix is given by
\be{\label{c12}}
C=\pmatrix{0&C_{10}\cr C_{10}&0}\,,
\ee
where $C_{10}$ is the ten dimensional $C_-$ for the matrices
$\gamma_0,\dots,\gamma_9$.

We are now interested in the compactification of the twelve
dimensional theory. We shall first discuss the effects of the
compactification on the spin spaces, and compare these with those for
string theory and $M$-theory two-branes. If the twelve dimensional theory is to relate to
these theories then it must produce the right spinors upon
compactification. 

\subsection{Projection of the spin space}

Recall that, in  order to define the
$(2+2)$-brane in twelve dimensions, we must work in the $\mbox{sio}_2$
superspace which must be defined in terms of the restricted spinors
satisfying (\ref{constraintp})
\be\label{constraint2}
\delta\theta_\alpha(\Gamma^{\mu\nu})^{\alpha\beta}\theta_\beta
M_{\mu\nu}=0\,.
\ee
If we wish not to restrict the possible rotation states, this may be
re-written as 
\be\label{matrixid}
\widetilde{{\cal{P}}}\Gamma^{\mu\nu}C^{-1}{{\cal{P}}}=0\,.
\ee
In order to assist the compactification to ten and eleven dimensions,
we  shall consider the cases
\ben\label{projectors}
{\cal{P}}_{10}&=&\frac{1}{2}(1+\Gamma_0\dots\Gamma_9)= 
\frac{1}{2}\pmatrix{I+J&0\cr 0&I+J}\,\mbox{ and }\nonumber \\
{\cal{P}}_{11}&=&\frac{1}{2}(1-\Gamma_0\dots\Gamma_9\Gamma_{11})=
\frac{1}{2}\pmatrix{I&J\cr J&I}\,.
\een
These two projections restrict the twelve dimensional spinors to be of the form
\be\label{weylspinor}
\psi_{11}=\pmatrix{\alpha\cr \delta\cr \alpha\cr -\delta}\hskip0.5cm \mbox{ and } \hskip0.5cm\psi_{{10}}=\pmatrix{\alpha\cr 0\cr \gamma \cr 0} \,,
\ee 
and both  ${\cal{P}}_{10}$ and ${\cal{P}}_{11}$ satisfy the equation
(\ref{constraint2}), provided we further project out an additional
half of the spinor degrees of freedom.  
To see this is the case for ${\cal{P}}_{10}$, we first note that
$\{\Gamma_0\dots\Gamma_9,\Gamma^{\mu\nu}C^{-1}\}=0$ for
$\mu,\nu=0\dots 9$ and $\mu,\nu=$ 11 
or 12. In these  situations, the matrix identity (\ref{matrixid})
is satisfied, since $(1+\Gamma_0\dots\Gamma_9)(1-\Gamma_0\dots\Gamma_9)=0$,
and we therefore are done. It remains to check the case for which
we have $\mu=0\dots 9$ and $\nu=11$ or 12. These  terms fail to
satisfy the equation in the same way, since
$(\Gamma^{\mu\nu})^{\alpha\beta}$ commutes with ${\cal{P}}_{10}$ if
one and only one of $\mu$ or $\nu$ takes value of 10 or
11. In order to overcome this problem we must further project out
another half of 
the spinor degrees of freedom. 
 Substituting the explicit
representation of the gamma matrices we 
find that the extra condition which needs to be satisfied is that
\be
\pmatrix{\alpha&0}C_{10}\gamma^\mu\pmatrix{\beta\cr 0}\,,
\ee
which requires an additional rank(${\frac{1}{2}}$)
projection on the spinors, restricting $\alpha$ and $\beta$ in
much the same way as a $\kappa$-symmetry would.

We shall now discuss the relation between the theory with these
projectors and $p$-brane
theories in lower dimensions by compactification. At first glance it
may seem very unlikely that the $(2+2)$-brane could possibly reduce down
to Green-Schwarz strings and membranes, due to the different types of
supersymmetry for these theories. As we shall see, however, this is
not so.

In order to proceed, recall that to perform  a
dimensional reduction in a 
direction defined by a vector $n$ with  a background 
metric $\eta_{\mu\nu}$,  we act on the space with projection operators
${h^\mu}_\nu={\eta^\mu}_\nu\pm n^\mu n_\nu$. We need to explicitly  define
analogous operators ${H^\alpha}_\beta$ to act on the spin space so
that we can deduce the nature of the lower dimensional spinors. 
 The
form of ${H^\alpha}_\beta$ will depend on whether we perform a timelike,
spacelike or null reduction. In the compactified spin space we define
a new set of Gamma matrices,
$\{\hat{\Gamma}_\mu\}$, as 
\be\label{gammanew}
\widehat{\Gamma}_\mu={{H}}\Gamma_\mu \tilde{{H}}\,,
\ee
in matrix notation. It is with respect to these lower dimensional
gamma matrices that we discuss the chirality of the compactified
spinors.

\subsubsection{Double Null Reduction}
We now define two vectors which are null with respect to the (10,2)
flat space metric (\ref{clifford});
$u=\frac{1}{\sqrt{2}}(0,\dots,0,1,1) \mbox{ and }
v=\frac{1}{\sqrt{2}}(0,\dots,0,1,-1)$, so that
$u^2=v^2=0$ and $u.v=+1$.  The spin space double null operators are
then given by 
\be\label{hnull}
{{(H_{D\pm})}^\alpha}_\beta=\frac{1}{2}\left({I^\alpha}_\beta\ \pm {(\Gamma^{\mu\nu}u_\mu
u_\nu)^\alpha}_\beta\right)=\pmatrix{I&0\cr 0 & 0}\mbox{ or }\pmatrix{0&0\cr 0 & I}\,,
\ee
after  we substitute the explicit representation of the gamma
matrices (\ref{gamma12}). 
Note that ${H_D}^2={H_D}$ so we have a true projection operator.
Since the $H_{D}$ is a double null projector, the action of this
operator produces spinors in a spacetime of signature $(9,1)$.  Using
the expression (\ref{gammanew}) we can 
evaluate the new ten dimensional gamma matrices, $\hat{\Gamma}$, in a
(reducible) 
64 dimensional  representation. As one would expect, $\hat{\Gamma}_p$
is non-zero if $p=0\dots 9$, and the other two gamma matrices are projected to
zero under the effect of ${H_D}$ yielding 
\be
\hat{\Gamma}_p=\pmatrix{\gamma_p&0\cr 0&0}
 \mbox{ or }\pmatrix{0&0\cr 0&\gamma_p} \hskip1cm p=0\dots 9\,.
\ee
To discover the effect of the compactification on the chirality of the
spinors we must  define
the new $\hat{\Gamma}_{11}$ matrix in the projected space
\be\label{gamma11new}
{\hat{\Gamma}}_{11}=\hat{\Gamma}_0\dots\hat{\Gamma}_9=
\pmatrix{J&0\cr 0&0} \mbox{ or }\pmatrix{0&0\cr 0&J}\,.
\ee
The compactified  spinors are now defined by the relationship
\be
\hat{\psi}={H_D}\psi\,.
\ee
We find that the projected spinor $\psi_{10}$,
(\ref{weylspinor}), becomes a pair of 16 component spinors of
the {\underline{same}} chirality, with respect to ${\hat{\Gamma}}_{11}$,  under
the action of the  $H_{D_\pm}$; we have thus obtained the  spectrum
of the type IIB string 
theory.

\subsubsection{Timelike reduction from twelve dimensions}
We now perform the reduction along the timelike direction
$t=(0,0,\dots,0,1), t^2=-1$. The relevant projection operators are given by
\be\label{ht}
{({H_{t\pm}})^\alpha}_\beta=\frac{1}{2}({I^\alpha}_\beta\pm ({{\Gamma^{13}\Gamma^\mu
t_\mu})^\alpha}_\beta)=\frac{1}{2}\pmatrix{I&\pm J\cr
\pm J&I}\,,
\ee
which happens to be of the same form as the projection operator
${\cal{P}}_{11}$. 
In this case the new eleven dimensional chirality projection matrix is  
given by 
\be
\hat{\Gamma}_{12}=(H_t\Gamma_0{\tilde{H}}_t)\dots(H_t\Gamma_{11}{\tilde{H}}_t)=\frac{1}{2}\pmatrix{0&\pm
J\cr
\pm J&0}\,.
\ee

Compactification of the  projected spinor $\psi_{11}$, gives one zero
spinor and  a
single  non-zero eleven dimensional
spinor of the form 
\be
{\hat\psi}_{11}=\pmatrix{\alpha\cr \delta\cr \alpha\cr -\delta}\,,
\ee
which has  32 real components, from the eleven dimensional
point of view.  These are the
correct degrees of freedom for the  M-theory two-brane. From this theory we may
further compactify on a circle or a $\zz_2$ orbifold of the circle to produce
the type IIA string or heterotic $E_8\otimes E_8$ string
respectively \cite{HW:}.

\subsection{Compactification of the Superspace}
Although the arguments in the previous section show that the spinor
degrees of freedom after compactification correspond to the IIB string
theory  and the M-theory, we have not shown that the
compactification yields the correct supersymmetric theory given by the
action  (\ref{brane}), which is invariant under the action
of the super \po group SIO$(S,T)_1$. For the twelve dimensional theory
to truly relate to lower dimensions we must clearly be able to produce
the usual type of supersymmetry in a natural way. The
superspace in twelve dimensions is effectively predetermined by the
requirements of the existence of a brane action, in that it must be
the superspace for  the supergroup SIO$(10,2)_2$. 
We now must investigate the effect that the compactification has on
this superspace. We shall specifically consider the effects of the double-null
compactification (the analysis is similar for
the timelike case).

 Recall that the basic action in twelve dimensions is
given by (\ref{superaction}),
\be\label{firstorder}
S=\int d^{4}\xi\, \mbox{det}[\Pi_i^A\Pi_j^B{\cal{G}}_{AB}]^{\frac{1}{2}}\,,
\ee
with
\be
{\cal{G}}=\pmatrix{\eta&0\cr 0&C}\,.
\ee
and $\Pi^A=(dX^\mu,d\theta^\alpha)$.
Since the fundamental geometric object of the theory is the
superspace, it is natural to consider the effects
of the compactification on the bosonic and fermionic indices together
in a superspace projection ${{\cal{H}}^A}_B$, for $A,B=(\mu,\alpha)$. 
Schematically, ${{\cal{H}}^A}_B$  can be written as
\be\label{superproj}
{\cal{H}}=\pmatrix{B&Y\cr Z&F}\,,
\ee
where $B\equiv \{{B^\mu}_\nu\}$ is the projector which acts on the
purely bosonic part of the superspace and $F\equiv \{{F^\alpha}_\beta\}$ is the
projector which acts on the spin space. For example, for the double
null reduction case we have the
${B^\mu}_\nu={h^\mu}_\nu={\eta^\mu}_\nu+\frac{1}{2}(u^\mu v_\nu - v^\mu u_\nu)$
and ${F^\alpha}_\beta = {H^\alpha}_\beta=\frac{1}{2}{(I+\Gamma^{\mu\nu}u_\mu
v_\nu)^\alpha}_\beta$.
For a general compactification with these fermionic and bosonic parts,
the matrices   $Y\equiv\{{Y^\mu}_\alpha\}$ and
$Z\equiv\{{Z^\alpha}_\mu\}$ may be arbitrarily chosen. When Cremmer
and Julia constructed the $SO(8)$ supergravity, \cite{cj:so8}, they
effectively  considered the
case for which the superspace projector was block diagonal. This is by
no means a necessary choice, and in pure bosonic Kaluza-Klein theories
would correspond to a trivial compactification. We shall for the
moment leave the  terms $Y$ and $Z$ arbitrary, and
investigate the effect of the compactification on the $(2+2)$-brane action. 

Under the action of the superspace projection, the superspace metric
schematically transforms as
\ben\label{newg}
{\cal{G}}\rightarrow \hat{\cal{G}}={\cal{HG}}\tilde{{\cal{H}}}&=&\pmatrix{B&Y\cr
Z&F}\pmatrix{\eta&0\cr
0&C}\pmatrix{\tilde{B}&-\tilde{Z}\cr\tilde{Y}&\tilde{F}}\nonumber \\
&=&\pmatrix{B\eta\tilde{B}+YC\tilde{Y}&-B\eta\tilde{Z}+YC\tilde{F}\cr
Z\eta\tilde{B}+FC\tilde{Y}&-Z\eta\tilde{Z}+FC\tilde{F}}\,,
\een
where the tilde denotes matrix transpose. The minus sign occurs for
the $\tilde{Z}$ in the transpose of ${\cal{H}}$ because we are dealing
with supermatrices,  as
opposed to ordinary matrices. 
We now make the restriction that the purely bosonic part of the
transformed superspace metric should equal the projection of $\eta_{\mu\nu}$ by
${B^\mu}_\nu$. We similarly require that the fermionic sector should be
the same as the projection of the charge conjugation matrix by the
spin projector ${H^\alpha}_\beta$. This leads to the constraints on Y
and Z
\ben\label{yzconstraints}
(YC\tilde{Y})^{\mu\nu}&\equiv& {Y^\mu}^\beta C_{\alpha\beta}
\tilde{Y}^{\beta\nu}=0\nonumber\\
(Z\eta\tilde{Z})^{\alpha\beta}&\equiv&
Z^{\alpha\mu}\eta_{\mu\nu}\tilde{Z}^{\nu\beta}=0\,.
\een
These requirements essentially mean that the squares of  $Z$ and $Y$ must be
antisymmetric in the bosonic indices and symmetric in the fermionic
indices respectively. Thus the equations (\ref{yzconstraints}) are satisfied if we
choose
\be\label{yz}
Y^{\mu\beta}=\tilde{Z}^{\mu\beta}=-\frac{1}{2}\theta_\alpha(\Gamma^\mu)^{\alpha\beta}
\,.
\ee
Or course, we could choose that $Y=Z=0$, but this would lead to
another trivial superspace in lower dimensions, not the $\mbox{sio}_1$
that we are seeking.  Note that we only
require that the two sides of the  equation (\ref{yz}) be  proportional to each
other, but we choose the factor of $-\frac{1}{2}$ for convenience. 
Finally we suppose that all the spacetime indices have been acted upon
by $B$ and all the spinorial indices by $F$. This is in effect what is
achieved in the compactification
\be
FY=BY=Y\,.
\ee
We may now calculate the effects of the transformation (\ref{newg}) on the
Lagrangian for the action (\ref{superaction})
\ben\label{newlag}
{\cal{L}}=\Pi^A {\cal{G}}_{AB}\Pi^B&\rightarrow& \hat{\cal{L}}=\Pi^A
{\tilde{\cal{H}}_A}^{\,\, B}{\cal{G}}_{BC}{{\cal{H}}^C}_D\Pi^D\nonumber\\
&=&\pmatrix{dX^\mu&d\theta^\alpha}\pmatrix{B_{\mu\nu}&0\cr 4\tilde{Y}_{\alpha\nu}&F_{\alpha\beta}}\pmatrix{dX^\nu\cr
d\theta^\beta}\nonumber\\
&=&dX^\mu\eta_{\mu\nu}dX^\nu-2dX^\mu\bar{\theta}\Gamma_\mu d\theta\,,
\een
where the coordinates now all lie in the compactified superspace. 
The beauty of this construction is that the action principle may now
be reformulated as
\be
S=\int d^p\xi\,
\mbox{det}[(\diff_iX^\mu-\bar{\theta}\Gamma^\mu \diff_i\theta)\eta_{\mu\nu}(\diff_jX^\nu-
\bar{\theta}\Gamma^\nu \diff_j\theta)]^{\frac{1}{2}}\,,
\ee
which is precisely the form of the $\mbox{sio}_1$ invariant action (\ref{brane}). 
Thus we have shown that by performing a natural  superspace projection on a 
trivial higher dimensional superspace, we reproduce the usual twisted
superspace  of
standard supersymmetry. 

\subsection{Compactification of the super-algebra}
We now turn to the question of the dimensional reduction of
the supersymmetry algebra, (\ref{projalg}), which was is so essential in the
construction of super $(2+2)$-brane. For the reduction of the $(2+2)$-brane 
consistently to yield the usual Minkowski strings and membranes we must  
retrieve the sio$(9,1)_1$ or sio$(10,1)_1$ algebras after compactification. 
The important term to discuss is the anticommutator of the spinor
generator with itself  for the sio$(10,2)_2$ algebra, after we have
taken the projection of the spin space, 
\be
\{{{\cal{P}}^\alpha}_{\hat{\alpha}} Q^{\hat{\alpha}},{{\cal{P}}^\beta}_{\hat{\beta}}Q^{\hat{\beta}}\}=\frac{1}{2}{{\cal{P}}^\alpha}_{\hat{\alpha}}(\Gamma^{\mu\nu})^{\hat{\alpha}\hat{\beta}}M_{\mu\nu}{{\cal{P}}^\beta}_{\hat{\beta}}\,.
\ee
The only terms which survive on the right hand side of the projected
anticommutator are those for which the matrix identity
(\ref{matrixid}) is not satisfied for the projectors ${\cal{P}}_{10}$
or ${\cal{P}}_{11}$.  This yields
\ben\label{ac}
\{({{{\cal{P}}_{10}})^\alpha}_{\hat{\alpha}}Q^{\hat{\alpha}},{({{\cal{P}}_{10}})^\beta}_{\hat{\beta}}Q^{\hat{\beta}}\}&=&\frac{1}{2}{{({\cal{P}}_{10})}^\alpha}_{\hat{\alpha}}\left((\Gamma^{\mu\,
11})^{\hat{\alpha}\hat{\beta}}M_{\mu\, 11}+({\Gamma^{\mu\,
12}})^{\hat{\alpha}\hat{\beta}}M_{\mu\,
12}\right){{({\cal{P}}_{10})}^\beta}_{\hat{\beta}}\nonumber \\
\{({{{\cal{P}}_{11}})^\alpha}_{\hat{\alpha}}Q^{\hat{\alpha}},{({{\cal{P}}_{11
}})^\beta}_{\hat{\beta}}Q^{\hat{\beta}}\}&=&\frac{1}{2}{{({\cal{P}}_{11})}^\alpha}_{\hat{\alpha}}\left((\Gamma^{\mu\,
12})^{\hat{\alpha}\hat{\beta}}M_{\mu\, 12}\right){{({\cal{P}}_{11})}^\beta}_{\hat{\beta}}\,.
\een
We now need to determine the form of the rotation generator,
$M_{\mu\nu}$. A realisation of the algebra sio$(S,T)_2$ is given by 
\be\label{mrep}
M_{\mu\nu}=X_{[\mu}\diff_{\nu]}+\frac{1}{2}\theta_\alpha\left(\Gamma_{\mu\nu}\right)^{\alpha\beta}\diff_\beta\,,
\ee
where we may choose
$Q_\alpha=\diff_\alpha=\frac{\diff}{\diff\theta^\alpha}$.
If we consider projecting the spinors by the projectors
${{\cal{P}}_{11}}$ and ${{\cal{P}}_{10}}$ then the second term in this
expression  vanishes identically, in which case we find that
\be
M_{\mu\nu}=X_{[\mu}\diff_{\nu]}\,.
\ee
We now perform the dimensional reduction on the anticommutators
(\ref{ac}), by acting with the operators ${{(H_{D\pm})}^\alpha}_\beta$
(\ref{hnull}). The double null reduction leads to 
\ben\label{reducedac}
\{{Q_{10}^{+\alpha}},{Q_{10}^{+\beta}}\}&=&({\gamma^{p}}^+)^{\alpha\beta}P_p\nonumber\\
\{{Q_{10}^{-\alpha}},{Q_{10}^{-\beta}}\}&=&({\gamma^{p}}^-)^{\alpha\beta}P_p\nonumber\\
\{{Q_{10}^{+\alpha}},{Q^{-\beta}_{10}}\}&=&0\nonumber\\
P_p&=&\frac{\diff}{\diff X^p}\,,
\een
where $Q^{\pm\alpha}_{10}$ is a positive/negative chirality Spin(9,1)
spinor, and ${\gamma_{p}}^\pm$ is a positive/negative projection of
the gamma matrix $\gamma_p$ for which $p=0\dots 9$.   
For the timelike compactification to eleven dimensions, one of the projectors
${{(H_{t\pm})}^\alpha}_\beta$ sends everything to zero, the action of
the other operator  gives 
\be{\label{reducedac2}}
\{{Q_{11}^{\alpha}},{Q_{11}^{\beta}}\}=(\hat{\gamma^p})^{\alpha\beta}P_p\,,
\ee
where ${\hat{\gamma}}^p$, for $p=1\dots 11$, are new gamma matrices
obtained by reduction of the twelve dimensional $\Gamma_p$, for
$p=1\dots 11$.
We thus see that the compactifications from the twelve dimensional
theory gives us the  correct basic $\mbox{sio}_1$ superalgebra that
we require. This is by virtue of the  superspace projection
(\ref{constraint})
required  to define the $\mbox{sio}_2$ $(2+2)$-brane action. It was
precisely this constraint which prevented the appearance of  ten or
eleven dimensional $(\gamma^{pq})^{\alpha\beta}M_{pq}$ terms in the
projected algebras (\ref{reducedac}) and (\ref{reducedac2}). Note that,
although ${(\Gamma^{12})^\alpha}_\beta$ is compactified to zero in
eleven dimensions, the term ${(\Gamma^{12p})^{\alpha\beta}}$ is not,
since the projection operator ${\cal{P}}_{11}$ does not commute with
the twelve dimensional charge conjugation matrix $C$. A similar
comment applies to the 
ten dimensional compactification. 

There is  now one final question to be asked: what does the full
twelve dimensional $\{Q,Q\}$ anticommutator, (\ref{a11}),  reduce to in ten and eleven
dimensions? The full set of the possible $\{Z^{(i)}\}$ is given by $\{Z^{(2)},
Z^{(3)},Z^{(6)},Z^{(7)},Z^{(10)},Z^{(11)}\}$. Since we are
compactifying over only one or two directions, the
compactification of the ${Z^{(i)}}$ with $i$ even  will
all be qualitatively the same; a similar statement holds for the
odd case. We thus need to determine the compactification of the
$\Gamma^{\mu_1\mu_2\mu_3}$ term. We find that for the double null
reduction, $(\Gamma^{\mu_1\mu_2\mu_3})^{\alpha\beta}$ is projected to
$(\Gamma^{p 11,12})^{\alpha\beta}$ and
$(\Gamma^{p_1p_2p_3})^{\alpha\beta}$. For the eleven dimensional case
we find that only the term ${\Gamma}^{12pq}$ survives the
projection. For the 7 and 11 index cases we need just add 4 or 8 $p$-type
indices to these expressions. If we then act with the compactification
operators, $H^\pm$, then  we find 
compactified versions of (\ref{a11}):
\ben\label{newalg}
\{{Q_{10}^{\pm\alpha}},{Q_{10}^{\pm\beta}}\}&=&({\gamma^{p}}^\pm)^{\alpha\beta}P_p+({\gamma^{
p_1\dots p_5\pm}})^{\alpha\beta}Z_{p_1\dots p_5}^\pm
+({\gamma^{
p_1\dots p_9\pm}})^{\alpha\beta}Z_{p_1\dots p_9}^\pm\nonumber \\
\{{Q_{10}^{\pm\alpha}},{Q_{10}^{\mp\beta}}\}&=&({\gamma^{p}}^\pm)^{\alpha\beta}\tilde{Z}_p+
({\gamma^{p_1\dots p_3\pm}})^{\alpha\beta}\hat{Z}_{p_1\dots p_3}^\pm
+({\gamma^{
p_1\dots p_5\pm}})^{\alpha\beta}\hat{Z}_{p_1\dots p_5}^\pm
\nonumber
\\
&\,&\hskip1cm+({\gamma^{
p_1\dots p_7\pm}})^{\alpha\beta}\tilde{Z}_{p_1\dots p_7}^\pm
+({\gamma^{
p_1\dots p_9\pm}})^{\alpha\beta}Z_{p_1\dots p_9}^\pm\nonumber \\
\{{Q_{11}^{\alpha}},{Q_{11}^{\beta}}\}&=&(\hat{\gamma}^p)^{\alpha\beta}P_p+
(\hat{\gamma}^{p_1p_2})^{\alpha\beta}Z_{p_1p_2}+
(\hat{\gamma}^{p_1\dots p_5})^{\alpha\beta}Z_{p_1\dots p_5}+
(\hat{\gamma}^{p_1\dots p_6})^{\alpha\beta}Z_{p_1\dots p_6}\nonumber\\
&\,&\hskip1cm
+(\hat{\gamma}^{p_1\dots p_9})^{\alpha\beta}Z_{p_1\dots p_9}+
(\hat{\gamma}^{p_1\dots p_{10}})^{\alpha\beta}Z_{p_1\dots p_{10}}\,,
\een
where the $\tilde{Z^{(i)}}$ are new set of objects which are
antisymmetric in the spacetime indices. These arise in the commutator
of two ten dimensional spinors of opposite chirality, and thus
correspond to  string theory central charges.  
We now make use of the algebraic equivalence between a $p$-form and a
$D-p$-form in these supersymmetry algebras to find that
\ben\label{newalg2}
\{{Q_{10}^{\pm\alpha}},{Q_{10}^{\pm\beta}}\}&=&({\gamma^{p}}^\pm)^{\alpha\beta}P_p+({\gamma^{
p_1\dots p_5\pm}})^{\alpha\beta}Z_{p_1\dots p_5}^\pm\nonumber \\
\{{Q_{10}^{\pm\alpha}},{Q_{10}^{\mp\beta}}\}&=&({\gamma^{p}}^\pm)^{\alpha\beta}\tilde{Z}_p+
({\gamma^{p_1\dots p_3\pm}})^{\alpha\beta}\tilde{Z}_{p_1\dots p_3}^\pm
+({\gamma^{
p_1\dots p_5\pm}})^{\alpha\beta}\tilde{Z}_{p_1\dots p_5}^\pm
\nonumber \\
\{{Q_{11}^{\alpha}},{Q_{11}^{\beta}}\}&=&(\hat{\gamma}^p)^{\alpha\beta}P_p+
(\hat{\gamma}^{p_1p_2})^{\alpha\beta}Z_{p_1p_2}+
(\hat{\gamma}^{p_1\dots p_5})^{\alpha\beta}Z_{p_1\dots p_5}\,.
\een
These supersymmetry algebras (\ref{newalg2}) are {\it{precisely}}
those for the IIB and IIA superstrings, and for $M$-theory, \cite{t:pbranedem}, completing our analysis of
the dimensional 
reduction of the sio$(10,2)$ invariant superalgebra.

\section{Conclusion}
\setcounter{equation}{0}
We have presented in this paper the construction of a covariant
supersymmetric brane with worldvolume signature (2,2). This brane
propagates in a twelve dimensional spacetime with two timelike
directions, and for consistency the $(2+2)$-brane must be defined in a new
type of flat superspace which has no torsion. In order that this
superspace be well defined, the spinors must satisfy an unusual
spinor identity. This constraint leads to supersymmetry on the
brane. We have shown that this $(2+2)$-brane can be related to the
M-theory 2-brane and the type IIB string theory in ten dimensions under
compactification, without the need to go to nine dimensions, or to use
thirteen dimensions, as in \cite{stheory}. This is a
success since these two theories have 
until now had no direct single origin.  Some speculation along the
lines of replacing the brane  by a (2,1) string to produce the IIB
string and M-theory  has been
made by Kutasov,  
Martinec and O'Loughlin \cite{km,kmo}. In this work it seems necessary
to use a $(2+2)$-brane, which 
 could well be related  to the super $(2+2)$-brane presented here.

It would be interesting to study the relationship between the theory
presented in this paper and F-theory, since both require the signature
of spacetime 
to be (10,2). F-theory has had some successes and provides a mechanism
for explaining S-duality in the IIB string theory. There are clearly
some conceptual difficulties with these theories, however, due to
there being two timelike directions. For example, the notion of a
brane propagating through the spacetime becomes unclear. It has been
suggested that a twelfth dimension is merely auxiliary and is simply a
clever tool used to discuss lower dimensional theories. It should be
stressed, however, that the $(2+2)$-brane is an essentially twelve
dimensional object, whose definition was independent of consideration
of theories in
lower dimensions: all the constraints arise from a discussion
of twelve dimensional supersymmetry. This could point towards
the existence of a real extra dimension.

It seems plausible that the formalism presented in this paper could be
related to AdS supergravity in eleven dimensions and the massive
string theories and supergravity in ten dimensions. This is because
the quantities 
\be
X^A{\cal{G}}_{AB}X^B=\mbox{constant}\,,
\ee
arise naturally for the supersymmetry discussed in this paper. This is
the supersymmetric version of the constraint used to describe AdS in
dimension (d,1) as a hypersurface in dimension (d+1,2).

Perhaps the most serious problem with any theory which involves a
spacetime 
dimension greater than eleven is that we are lacking a local field
theory. This is not to say that such a theory does not exist, but 
merely that all the obvious constructions fail. If we suppose that the
twelve dimensional field theory should somehow be related to branes as in ten
and eleven dimensions, then the failure of the
standard methods to provide an answer  should  not be a
surprise, due to the very different nature of the $(2+2)$-brane 
to the traditional $p$-branes.
For the time being, the
resolution of this 
problem remains a mystery.

\appendix

\setcounter{equation}{0}
\section{Subalgebras of the de Sitter algebra}

We shall investigate the subalgebras of $N=1$ gradings of the  de Sitter
algebra, which is given in (\ref{desitter}). We include the anticommutator term\be\label{a11}
\{Q^\alpha,Q^\beta\}=\sum_{\mbox{\tiny{symmetric }}\Gamma}\frac{1}{k!}(\Gamma_{\mu_1\dots\mu_k})^{\alpha\beta}Z^{\mu_1\dots\mu_k}\,.
\ee
The full graded algebra is generated by the basic terms $\{P,M,Q\}$ and a set
of $Z^{(i)}$. Van Holten and Van Proeyen, \cite{t3},
checked the consistency of these algebras for given $Z^{(i)}$. They
found that, if we  identify $Z^{1}$ and $Z^{2}$ with $P$
and $M$ respectively, in the infinite radius limit  we have the following
terms in the algebra:
\be\label{a12}
[Z^i,Z^j]\sim 2y\left(\frac{i+j-k}{2}\right)!\left(\frac{i-j+k}{2}\right)!\left(\frac{-i+j+k}{2}\right)!Z^k\,.
\ee
In this equation we have suppressed the indices: it is only the form
of these commutators which is important in this discussion. 

We start from a maximal set of the $Z^{(i)}$ and search for
subalgebras. In twelve dimensions, for $C_-$  we have
$\{Z^1,Z^2,Z^5,Z^6,Z^9,Z^{10}\}$ as the possible generators,  and for
$C_+$ we have $\{Z^2,Z^3,Z^6,Z^7,Z^{10},Z^{11}\}$. We now present some
of the commutation relations   between the $\{Z^{(i)}\}$. We always have that 
\be
\left[Z^{2},Z^{(i)}\right]\sim Z^{(i)}\,,
\ee
and  for
the two different choices of the charge conjugation matrix we
have an additional   set of commutators. For the $C=C_-$ case we have 
\ben
\left[Z^{1},Z^{1}\right]&\sim& yZ^{2}\nonumber\\
\left[Z^1,Z^5\right]&\sim& yZ^{6}\nonumber\\
\left[Z^{1},Z^{6}\right]&\sim& yZ^{5}\nonumber\\
\left[Z^{1},Z^{9}\right]&\sim& yZ^{10}\nonumber\\
\left[Z^{1},Z^{10}\right]&\sim& yZ^{10}\nonumber\\
\left[Z^{5},Z^{6}\right]&\sim& y(Z^{1}+Z^{5}+Z^9)\nonumber\\
\left[Z^{5},Z^{9}\right]&\sim& y(Z^6+Z^{10})\nonumber\\
\left[Z^{5},Z^{10}\right]&\sim& y(Z^{5}+Z^{9})\nonumber\\
\left[Z^{6},Z^{9}\right]&\sim& y(Z^3+Z^9)\nonumber \\
\left[Z^{6},Z^{10}\right]&\sim& y(Z^6+Z^{10})\nonumber \\
\left[Z^{9},Z^{10}\right]&\sim& y(Z^{1}+Z^5+Z^9)\nonumber \\
\left[Z^{i},Z^{i}\right]&=&y(Z^2+Z^{6}+Z^{10})\mbox{ for } i\geq 5\,.
\een
From these terms it is clear that the only subalgebras of the algebra
generated by the $\{Z^{i}\}$, without setting some of the generators to
zero, is generated by $\{Z^{2},Z^{1}\}$ This subalgebra is the
same as the \po algebra, (\ref{poincare}), in the infinite radius
limit,  if we make the scaling
\be
Z^1=\left(\frac{2y}{m} \right)P,\hskip1cm Z^{2}=2yM\,.
\ee
For the $C=C_+$ case we find that 
\ben
\left[Z^{3},Z^{6}\right]&\sim& y(Z^{3}+Z^{7})\nonumber\\
\left[Z^{3},Z^{7}\right]&\sim& y(Z^{6}+Z^{10})\nonumber\\
\left[Z^{3},Z^{10}\right]&\sim& y(Z^{7}+Z^{11})\nonumber\\
\left[Z^{3},Z^{11}\right]&\sim& yZ^{10}\nonumber\\
\left[Z^{6},Z^{7}\right]&\sim& y(Z^{3}+Z^{7}+Z^{11})\nonumber\\
\left[Z^{6},Z^{10}\right]&\sim& y(Z^{6}+Z^{10})\nonumber\\
\left[Z^{6},Z^{11}\right]&\sim& y(Z^{7}+Z^{11})\nonumber\\
\left[Z^{7},Z^{10}\right]&\sim& y(Z^{3}+Z^{7}+Z^{11})\nonumber\\
\left[Z^{7},Z^{11}\right]&\sim& y(Z^{6}+Z^{10})\nonumber\\
\left[Z^{10},Z^{11}\right]&\sim& y(Z^{3}+Z^{7}+Z^{11})\nonumber\\
\left[Z^3,Z^3\right]&\sim& y(Z^2+Z^6)\nonumber\\
\left[Z^{i},Z^{i}\right]&\sim&y(Z^2+Z^{6}+Z^{10})\mbox{ for } i\geq 6\,
\een
In this case we also find that there is only one subalgebra, which is
generated by   $\{Z^2,Z^6,Z^{10}\}$. This is algebraically equivalent
to the algebra generated by  $\{Z^2,Z^6\}$ since in twelve dimensions
$Z^2$ has the same degrees of freedom as $Z^{10}$. 

\bigskip
Note that for any given algebra, when we take the \po limit we must set y to be
zero, in which  case all of the $Z^{(i)}, \, i>2$ terms become central
charges. For gradings of the \po algebra, any of the $Z^{(i)}$ can be
introduced: those sets which form subalgebras of the full graded de Sitter
algebra  have a natural origin.

\section{}

We wish to find solutions of the identity (\ref{formeq}). 
Since the $d\theta$ terms are real and commuting, all that is
required is to 
show that
\be\label{a1}
(\Gamma_\mu{\cal{P}})_{(\alpha\beta}(\Gamma^{\mu\nu\rho}{\cal{P}})_{\gamma\delta)}=0\,,
\ee
where the round brackets denote symmetrisation of indices.
If the projection operator is the identity, then the equation is
automatically satisfied, since $\Gamma^\mu C$ or $\Gamma^{\mu\nu\rho}C $
is antisymmetric. 

We now look at the situation for which ${\cal{P}}$ is not the
identity. We shall use a basis of the Clifford algebra for which
$C$ anticommutes or commutes with $\Gamma^\mu$ for $\mu$ timelike or
spacelike respectively. We shall consider signature $(10,2)$. 

Firstly, if we choose ${\cal{P}}$ to be the Weyl projector,
${\cal{P}}=\frac{1}{2}(1+\Gamma_1\dots\Gamma_{12})$, then the identity
(\ref{a1}) is clearly satisfied,  since one of the two terms in the
expansion is 
antisymmetric in the spinor indices.

We shall now check the  identity for another  projector which is of use
to us: 
${\cal{P}}_{11}=\frac{1}{2}(1+X)$, where $X=\Gamma_{\mu_1}\dots\Gamma_{\mu_{11}}$,
for which exactly one of the $\mu_1,\dots,\mu_{11}$ is a timelike
index. We shall use $C_+$ as the charge conjugation matrix, so that
$\Gamma^\mu C$ is antisymmetric in the spinor indices. 

The identity (\ref{a1}) becomes
\be\label{a3}
(\Gamma_a
C{\cal{P}})_{(\alpha\beta}(\Gamma^{a\nu\rho}{\cal{P}})_{\gamma\delta)}
+(\Gamma_{12}
C{\cal{P}})_{(\alpha\beta}(\Gamma^{12\nu\rho}{\cal{P}})_{\gamma\delta)}
=0\,,
\ee
where $a$ takes values from 1 to 11. We now note that   $(\Gamma_{12}
CX)$ and $(\Gamma^{abc}CX)$ are antisymmetric 
in the spinor indices, whereas $(\Gamma^{12\nu\rho}CX)$ and  $(\Gamma_{a} CX)$are symmetric in
the spinor indices, thus the identity (\ref{a1}) reduces to 
\be
(\Gamma_{a}
CX)_{(\alpha\beta}(\Gamma^{a12b}{\cal{P}})_{\gamma\delta)}=0\,,
\ee
where $a,b$ run over $1,\dots,11$. 

Substituting in the  particular representation (\ref{gamma12}) of the gamma matrices, we
see that this is equation is precisely the matching condition
equation, (\ref{formeq}), 
for a 2-brane moving in eleven dimensional Minkowski spacetime, for
which we know that the equation is satisfied. Thus we are done.

A similar procedure applies to case where the projector is ${\cal{P}}=\frac{1}{2}(1+\Gamma_0\dots\Gamma_9)$.

\end{document}